\documentclass[preprint,12pt]{aastex}

\newcommand{\Msun}{M_{\sun}}

\newcommand{\second}{{\mathrm{s}}}

\newcommand{\ice}{{ICECUBE }}

\def\ltaprx {\lower .1ex\hbox{\rlap{\raise .6ex\hbox{\hskip .3ex
        {\ifmmode{\scriptscriptstyle <}\else
                {$\scriptscriptstyle <$}\fi}}}
        \kern -.4ex{\ifmmode{\scriptscriptstyle \sim}\else
                {$\scriptscriptstyle\sim$}\fi}}}
\def\gtaprx {\lower .1ex\hbox{\rlap{\raise .6ex\hbox{\hskip .3ex
        {\ifmmode{\scriptscriptstyle >}\else
                {$\scriptscriptstyle >$}\fi}}}
        \kern -.4ex{\ifmmode{\scriptscriptstyle \sim}\else
                {$\scriptscriptstyle\sim$}\fi}}}

\shorttitle{Neutrinos from Collapsars}
\shorttitle{Pruet, J.}

\begin{document}

\title{Neutrinos from the Propagation of a Relativistic Jet Through a Star}
\author{Jason Pruet}
\affil{
N-Division,
    Lawrence Livermore National Laboratory,
    Livermore CA 94550
}

\begin{abstract}

We discuss the neutrino signature of a relativistic jet
propagating through a stellar envelope, a scenario realized in the
collapsar model for Gamma Ray Bursts (GRBs). It is shown that the
dramatic slowing of the jet deep within the star is accompanied by
inelastic neutron-nucleon collisions and the conversion of a
substantial fraction of the jet kinetic energy to neutrinos. These
neutrinos have observed energies in the range two to tens of GeV and
an estimated detection rate comparable to or larger than the detection
rate of GeV neutrinos from other GRB-related processes. The time delay
between the arrival of these neutrinos and the GRB photons is tens of
seconds. An observation of this delay would provide an indication that
the GRB jet originated in a massive star.

\end{abstract}


\section{INTRODUCTION}

In this paper we discuss neutrino production in collapsars
\citep{woo93, mac99}. Collapsars, or failed supernovae, are one of two
or a few currently favored explanations for the origin of long
duration GRBs. In collapsars, the relativistic jet which shocks and
ultimately gives rise to observed GRB photons must first pass through
a massive stellar mantle. Here it is shown that this passage is
associated with an interesting neutrino signal.

A number of GRB-related neutrino sources have been previously investigated.
These sources can be roughly divided according to whether they involve
pion production via photomeson interactions or via strong interactions.
Neutrinos resulting from photomeson interactions typically have very 
high energies ($>10^2{\rm TeV}$) and are possibly the most easily detectable
in planned km-scale neutrino detectors such as \ice \citep{hal99}.
\cite{wax97} discussed photomeson production of neutrinos in collisionless
GRB shocks and argued that detection of these neutrinos would give information
on the spectra of photons and protons within the shocks, as well as place
stringent constraints on fundamental neutrino properties.

Neutrinos from inelastic nuclear collisions typically have energies of
the order of 10 GeV. Unless the phototube density in \ice and other next 
generation neutrino telescopes is higher
than currently planned, these detectors will have a small effective detector
area for such low energy neutrinos and will not be sensitive to them
\citep{bigice}. However, there is a growing indication that, like
neutrinos from ${\rm p-\gamma}$ interactions, these neutrinos may be
an important clue to conditions in and around the GRB central
engine.  \cite{bah00} discussed the neutrino signal from inelastic
collisions occurring during the dynamic decoupling of neutrons in the
acceleration stage of the evolution of a GRB fireball. These neutrinos
offer an indication of the composition and Lorentz factor of the GRB
jet. Neutron diffusion in GRB environments was considered in a more
general context by \cite{mes00}.  Those authors showed that internal
shocks at radii $r\sim 3\cdot 10^{11}{\rm cm}$ and transverse
diffusion of neutrons into the jet at much larger radii can lead to an
appreciable neutrino signal. The possibility of using neutrinos to
distinguish between collapsars and supranovae \citep{vie98,vie99} was
discussed by \cite{gue02}.  The scenario envisaged by Guetta and
Granot is unique in that it predicts TeV neutrinos from nuclear
processes. This is because of the presence in their model of protons
accelerated to very high energies in a pulsar wind bubble.

The multi-GeV collapsar neutrino signal we study here is related to
the mechanism giving rise to the variation in Lorentz factor of the
outgoing jet. As with any GRB model invoking internal shocks, the jet
variability must be substantial in order to explain observed temporal
variability in GRBs, as well as to allow efficient conversion of
kinetic energy to electromagnetic energy (see \cite{pir99} for a
general overview). In most GRB models, the jet variability is related
to a variability in the mass ablation rate or luminosity of the
central source driving the outflow.  By contrast, the jet variability
in the collapsar can arise from instabilities along the jet-stellar wall
interface \citep{zha02}. These instabilities operate at radii large
compared to the size of the central black hole. Consequently, the
instabilities act to slow a jet which has already had ample time to
reach ultra-relativistic velocities.  We point out that when free
neutrons are present in the jet, this slowing is not elastic and is
accompanied by a sizeable neutrino flux.

In the next section we discuss details of how the slowing of the
neutrons in the jet leads to the conversion of jet energy to
potentially observable neutrinos. Estimates of neutrino detection
rates and energies are presented. We will show that the neutrino
signal of the slowing of the jet can originate from radii some two
orders of magnitude smaller than the signal of processes discussed by
\cite{mes00}. Also, the neutrino production we discuss here is
relatively insensitive to the bulk Lorentz factor of the outgoing
jet, unlike other proposed sources of multi-GeV GRB
neutrinos.

\section{Pion and Neutrino Production Deep Within the Collapsar}

\begin{figure}
\epsscale{1.0}
\plotone{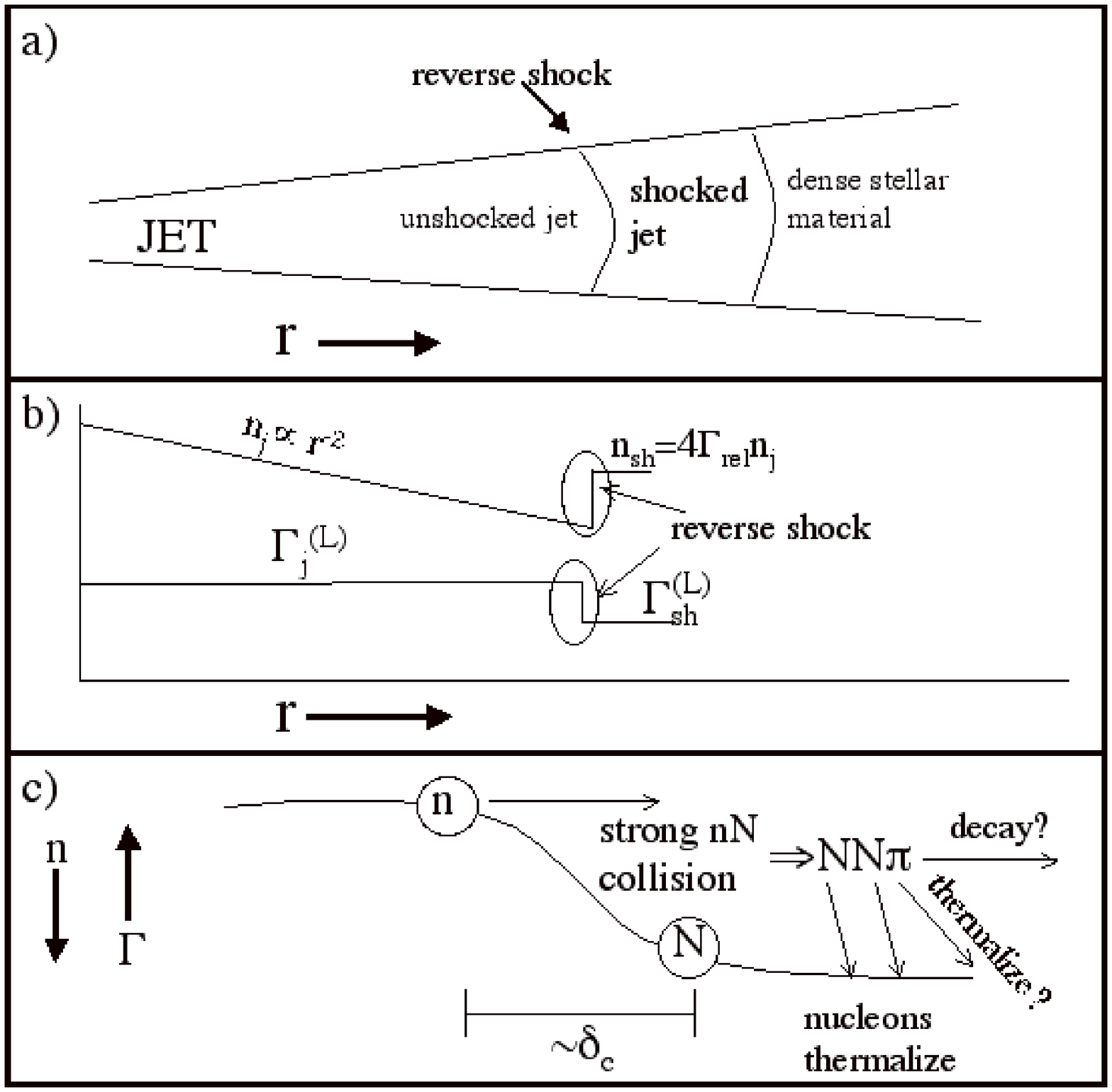}
\caption{ Illustration of the jet shocking and shock structure. Panel
a) shows the shocking of the jet propagating through the stellar
envelope. In this paper we are concerned with shocks occuring between
$r\sim 10^9-10^{11}{\rm cm}$. Panel b) shows the the baryon number
density (n) and jet Lorentz factor ($\Gamma$) in the un-shocked
(subscript $j$) and shocked (subscript $sh$) regions. Panel c) shows a 
close-up of the reverse shock (circled region in panel b). In panel
c) the shock width is $\delta_c$. This is the typical length over which 
the neutron-nucleon collisions slowing the jet occur. Neutrons will eventually
thermalize with the plasma in the shocked jet region. Depending on the details
of the shocking, pions created in inelastic neutron-nucleon collisions may or
may not thermalize before decaying.
\label{blob2}}
\end{figure}

The mechanism giving rise to the variability in the outgoing jet can be crudely
represented as the closing of a dense door composed of stellar material at
some point along the jet axis. This results in a forward shock propagating
into and accelerating the stellar material, and a reverse shock
propagating backwards and slowing the jet. Because the stellar material
blocking the passage of the jet is typically very dense relative to
the jet material, the forward shock will only be mildly relativistic and
uninteresting as far as a detectable neutrino signature is concerned.
The top panel of Fig. \ref{blob2} illustrates the shocking and structure
of the outgoing jet.

As particles in the jet traverse the reverse shock, they are
decelerated from a Lorentz factor $\Gamma_j^{(L)}$ to a Lorentz factor
$\Gamma_{sh}^{(L)}<\Gamma_j^{(L)}$. Here the superscript $(L)$ denotes
a quantity as measured in the rest frame of the collapsing star. The
relative Lorentz factor across the shock is $\Gamma_{rel}\approx
\Gamma_j^{(L)}/2\Gamma_{sh}^{(L)}$. Numerical simulations of
relativistic jets propagating in stars indicate that the typical
relative Lorentz factor is substantial. Zhang et al. studied the
propagation of a relativistic jet in a 15$\Msun$ star for 3 different
initial jet conditions. Typical values of $\Gamma_{rel}$ for their
simulations are given in Table \ref{thetable}.
The simulations of Zhang et al., as well as observations
of many peaks in observed GRB lightcurves, indicate that a substantial
fraction (half or more) of the jet undergoes a slowing via relativistic
shocks.

The gross properties of the reverse shock slowing the jet are given by
the Rankine-Hugoniot jump conditions. In particular, the proper baryon
number density in the shocked jet (i.e. the baryon number density as
measured in a frame comoving with the shocked fluid) is $n_{sh}\approx
4\Gamma_{rel} n_j$, where $n_j$ is the proper baryon number density in
the unshocked jet. For radii larger than $\approx 10^6\Gamma_j^{(L)}{\rm
cm}$, which is the region of interest in the present work, the
outgoing jet is coasting rather than accelerating \citep{coast}. In the
coasting regime, $n_j$ is determined by baryon number conservation as
$n_j=L/4\pi r^2 m_N c^3 (\Gamma_j^{(L)})^2$. Observations of GRBs
suggest that the isotropic equivalent luminosity $L$ is of order
$L\gtrsim 10^{52}{\rm erg/sec}$ \citep{frail}, and that typically
$\Gamma_j^{(L)}\gtrsim 200$ \citep{lith}.  The energy density in the
shocked jet fluid is $U_{sh}\approx \Gamma_{rel} n_{sh} m_N c^2$,
provided that the specific enthalpy in the unshocked jet is small
(i.e. that the jet is in the coasting regime). In the middle panel of
Fig. \ref{blob2} we show the evolution of $n_j$ and $\Gamma_j^{(L)}$
in the jet.

Pion production and associated neutrino production in the reverse
shock depend in detail on the processes mediating the shock.  It is
convenient in discussing the shock structure to consider separately
those particles which interact electromagnetically
(protons-$e^{\pm}$-photons) and the neutrons. Neutrons can only be
slowed by strong collisions with other nucleons. Because the length
scales associated with electromagnetic processes are small compared to
the neutron mean free path, an upper limit to the width of the shock
can be obtained by considering the case where strong nucleon-nucleon
collisions alone mediate the shock. Such collisional shocks have been
studied by \cite{mot51} in the non-relativistic case and by
\cite{cha79} and \cite{cer88} in the relativistic case.

 We take as an upper limit to the shock width the collisional width
$\delta_c \approx 1.9/\Gamma_{rel} n_j \sigma_{NN}$ found from the
Monte-Carlo simulations of \cite{cha79}. Here $\sigma_{NN}$ is the
strong scattering cross section and the shock width is as measured in
a frame comoving with the shock.  For center of mass energies a few
hundred MeV above the threshold for pion production, $\sigma_{\rm NN}
\approx 4\cdot 10^{-26} {\mathrm cm^2}$, and approximately $80\%$ of
this cross section goes towards inelastic production of pions (see
\cite{pdg} for a compilation of strong cross sections). For this work,
the important properties of $\delta_c$ are I) that because the shock
width is at most a few collisional mean free paths, the
neutron-nucleon collisions slowing the shock are hard and produce
pions, and II) because $\delta_c \propto 1/n_{sh} \sigma_{NN}$, the
number of inelastic collisions suffered by a typical neutron in
traversing the shock is approximately independent of $\Gamma_{rel}$. A
close-up view of inelastic neutron-nucleon collisions and associated
pion production in the reverse shock is shown in the bottom panel of
Fig. \ref{blob2}.

Protons in the shock will be slowed by electromagnetic, rather than
strong, processes. Deep within the star, where the shocks considered in
this paper are occurring, the plasma is optically thick and the thermalization
time for radiation is short. Consequently, we approximate the protons as
being in thermal equilibrium with the background $e^{\pm}/\gamma$ 
plasma\footnote{The opposite limiting case for the proton distribution
is that where the protons are Fermi-accelerated and achieve a power law
distribution. In this case, the maximum proton energy (typically much larger
than the proton rest mass) is that for which the acceleration and cooling
times are equal \citep{wax95}.}. The temperature of the protons in this
case is small, $kT\ll m_p c^2$, and the protons are non-relativistic in 
the shocked plasma frame.

Our assumption about the proton distribution function is in accord with
results from studies of shock propagation in type II SNe (e.g. \cite{ens92}).
However, \cite{mes01} have argued that protons
shocked in mildly relativistic collisions at large radii inside the
collapsar will be driven to a power law distribution, but that protons
shocked in the strongly relativistic collision between the jet head
and the stellar mantle may or may not have a power law
distribution. The question of the precise radius (presumably larger
than $10^{11}{\rm cm}$) at which shocks are collisionless rather than
radiative is not addressed here. As we are most interested in processes
at $r\lesssim 10^{11}{\rm cm}$, this issue is not so important for the
present study.

In the inelastic collisions slowing the neutrons, the total energy
going into pion production in the center of mass frame of the
collision is $E_{\pi}^{tot} \approx(1/2)(E_{cm}-2m_Nc^2)$
(e.g. \cite{man94}). Here $E_{cm}$ is the energy of the collision
in the center of mass frame, and the above equation is valid provided
that $E_{cm}-2m_Nc^2$ is larger than a few times the pion rest mass
energy. Very near the threshold for pion production, $E_{\pi}^{tot}\approx
E_{cm}-2m_Nc^2$. The average energy of individual pions created in the
collision is $E_{\pi}^{tot}/\xi$, where $\xi$ is the multiplicity of
produced pions. As measured in a frame at rest with respect to the star,
the energy of pions produced in a neutron-nucleon collision is
\begin{equation}
\label{epil}
E_{\pi}^{(L)}=\Gamma_{cm}E_{\pi}^{(tot)}=\left(m_N c^2\over 2\right)
\left( \Gamma_j^{(L)}+\Gamma_{sh}^{(L)}-2\Gamma_{cm}\right).
\end{equation}
Here we have made the assumption that the typical collision occurs
between a nucleon with Lorentz factor $\Gamma_j^{(L)}$ and a nucleon with
Lorentz factor $\Gamma_{sh}^{(L)}$.  In Eq. \ref{epil}, $\Gamma_{cm}$ is the
Lorentz factor of the center of mass frame as measured in the lab
frame. In terms of the Lorentz factors of the unshocked and shocked
portions of the jet, $\Gamma_{cm}\approx
\sqrt{\Gamma_{j}^{(L)}\Gamma_{sh}^{(L)}}$ if $\Gamma_{sh}$ is larger than
two or three, and $\Gamma_{cm}\approx \sqrt{\Gamma_j^{(L)} /2}$ if
$\Gamma_{sh}\approx 1$. The observed energy of an individual pion
created in a collision will again be a factor of $1/\xi$ smaller than
the estimate in Eq. \ref{epil}. For example, for $\Gamma_j^{(L)}=100$,
$E_{\pi}^{(L_)}/\xi \approx 10$, 12, and 7 GeV, for $\Gamma_{sh}=30$, 10,
and 1 respectively.

The neutrino signal of the inelastic collisions depends on how the
spectrum of pions produced in these collisions is modified by cooling
processes.  Pion cooling may proceed through inelastic strong
scatterings, or via electromagnetic interactions - coulomb scattering
off the background electrons, inverse compton scattering, and
synchrotron emission.

Cooling via strong scatterings is simply characterized by the
$\pi-$nucleon interaction time. We can roughly approximate $\pi-{\rm N}$
collisions as occurring in the shocked region (as opposed to the 
region of width $\delta_c$ characterizing the shock). In this case the
$\pi-$nucleon interaction time is
\begin{equation}
\label{taueqn}
\tau_{\pi-N}={1 \over c \sigma_{N\pi}n_{sh}},
\end{equation}
where $\sigma_{N\pi}$ is the nucleon-pion scattering cross section.
For center of mass energies a few hundred MeV larger than the pion
rest mass, $\sigma_{\rm N\pi}\approx 2.5\cdot 10^{-26}{\rm cm^2}$
\citep{pdg}. As with NN collisions, the bulk of this cross section
goes toward inelastic $\pi$ production. Though inelastic $\pi{\rm N}$
collisions may roughly conserve the total energy in pions, they tend
to reduce the average pion energy, making detection of the product
neutrinos more difficult.  Note that because $\sigma_{\rm N\pi}$ and
$\sigma_{\rm NN}$ are comparable, a more accurate expression for
$\tau_{\pi-N}$ should involve an effective nucleon number density
within the region of width $\delta_c$ (rather than the $n_{sh}$
appearing in Eq. \ref{taueqn}). Because the number density within the
shock is smaller than $n_{sh}$, Eq. \ref{taueqn} represents a lower
bound on $\tau_{\pi-N}$ (or an upper limit on the influence of $\pi N$
scatterings).

Pion energy loss via coulomb scattering off the background electrons
is given by the Bethe-Bloch formula for ionization losses (see \cite{groom}),
with the ionization potential for completely ionized material equal to
$\hbar \omega_p$. Here $\omega_p=\sqrt{4\pi n_e e^2/m_e}$ is the electron
plasma frequency, and $n_e$ is the electron number density. Neglecting
an unimportant (for our applications) logarithmic dependence on $n_e$,
pion energy loss via coulomb scattering is given by $dE/dx\approx
(2/3) \cdot 10^{-26}n_e {\rm GeV /cm}$.  The timescale for energy loss
via $\pi-e^{-}$ scatterings is then $\tau_{\pi-e^{-}}\approx
(\Gamma_{\pi}/2)(n_{\rm sh}/n_e) \tau_{\pi-N}$, with $\Gamma_{\pi}$
the pion Lorentz factor as measured in the shocked fluid frame. When
strong scatterings are inefficient at cooling the pions, the
temperature in the shocked plasma is generally too low for efficient
pair $e^{\pm}$ production. In this case $n_e \lesssim n_{sh}$ and
$\pi-e^{-}$ interactions are at most only as important as strong
scatterings. Pion energy loss via bremsstrahlung and pair production is
not important for pion energies less than about 100 GeV \citep{groom}.

Pions lose energy to the background photon and magnetic fields at a
rate $(dE_{\pi}/dt)_{\pi-\gamma}\approx
(c\Gamma_{\pi}^2)(m_e/m_{\pi})^2\sigma_{T}U_{em}$, where $\sigma_{T}$
is the Thomson cross section, $m_{\pi}$ is the pion mass, and $U_{em}$
is the energy density in the photon and magnetic fields. A cooling time 
$\tau_{\pi-\gamma}$ can be defined as 
$\tau_{\pi-\gamma}=E_{\pi}(dE_{\pi}/dt)_{\pi-\gamma}^{-1}$. Cooling of
pions via these electromagnetic interactions dominates over cooling
via strong interactions when
\begin{equation}
\left( \Gamma_{\pi}
\over 400 \right) \left( U_{em} \over n_{sh} m_N c^2 \right) \approx \left( {
\Gamma_{rel} \Gamma_{\pi} \over 400}\right) <1.
\end{equation}
Here we have made use of the fact that $U_{em}\sim U_{sh}$ for
relativistic shocks in an optically thick flow. Only in some of the
more extreme cases that we consider here is the above inequality not
satisfied. By contrast, for shocks at large radii ($r\sim 10^{12}{\rm
cm}$) when the protons are accelerated to high energies via plasma
processes, typical pions produced in photomeson
interactions have Lorentz factors of $\sim 10^6$, and synchrotron
cooling is more important than strong scattering.

In the shocked plasma rest frame the lifetime of a charged pion is
$\tau_{\pi-D}= 2.6\cdot 10^{-8} \Gamma_{\pi}\second$. The fraction of pions
that decay before suffering a strong scattering is 
$1/(\tau_{\pi-D}/\tau_{\pi-N}+1)$. The pion Lorentz factor relative to the
shocked plasma for which $\tau_{\pi-D}=\tau_{\pi-N}$ is
\begin{equation}
\label{picool}
\Gamma_{\pi,cool} \equiv  {7\cdot 10^2}{(r/10^{10}{\rm cm})^2
{(\Gamma_j^{(L)}/ 100)^2} \over
\Gamma_{rel} L_{52}}.
\end{equation}
Pions with a Lorentz factor greater than $\Gamma_{\pi,cool}$ will on
average scatter before decaying, and pions with a Lorentz factor
smaller than $\Gamma_{\pi,cool}$ will on average decay before
scattering.  Eq. \ref{picool} indicates that for $r\gtrsim 10^9{\rm
cm}$, pions produced in mildly relativistic collisions
($\Gamma_{rel}\sim 2$) will decay before cooling.  Conversely, for
$r\sim 10^9{\rm cm}$, pions produced in strongly relativistic shocks
($\Gamma_{rel}\gtrsim 10$, as in case JA in the calculations of Zhang
et al.) will cool before decaying. Neutrinos from the decay of cooled
pions will typically have observed energies $\lesssim 1{\rm GeV}$, and
will be difficult to detect. For $r\gtrsim 10^{10}{\rm cm}$,
Eq. \ref{picool} indicates that pions produced even in very
relativistic shocks will decay before cooling.

\begin{figure}
\epsscale{1.0}
\plotone{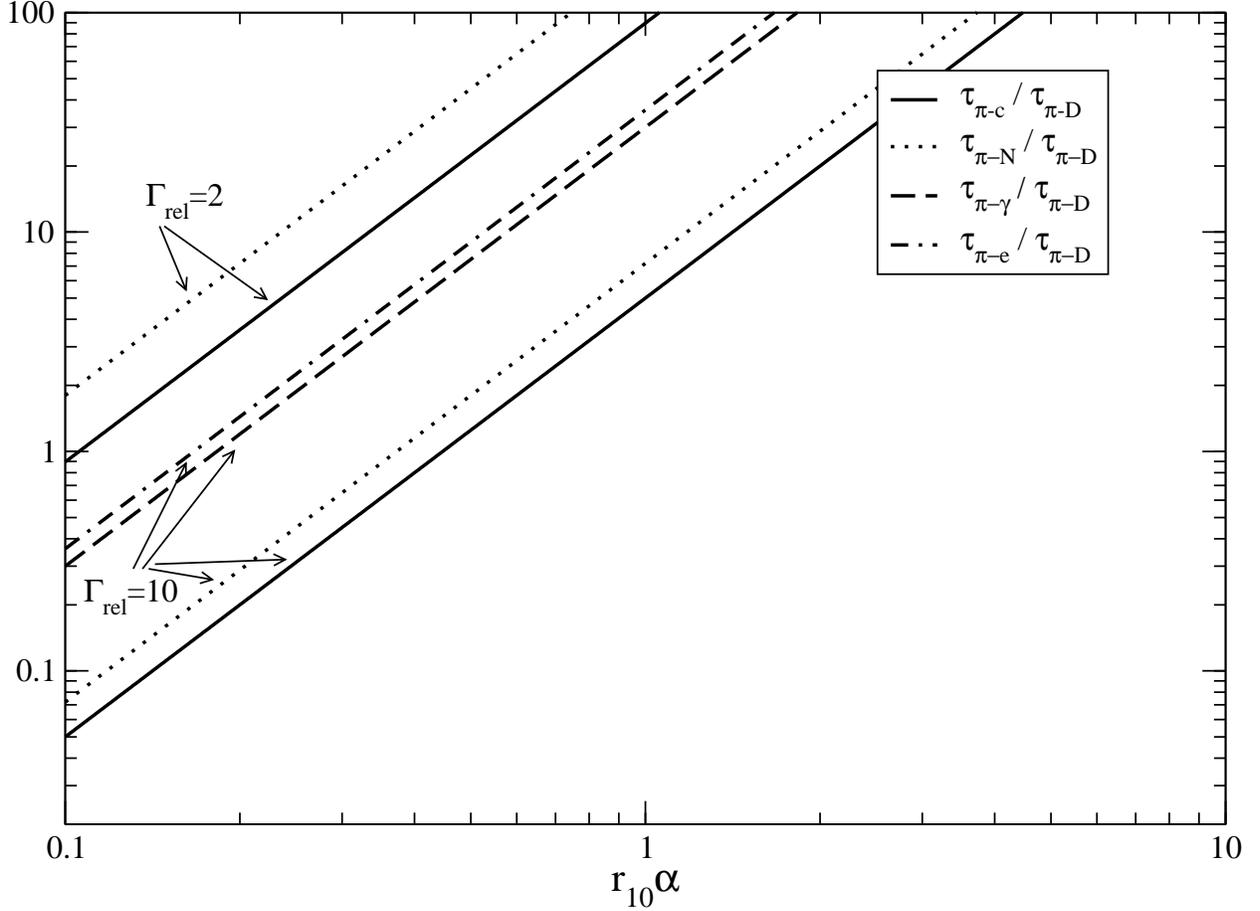}
\caption{Illustration of the ratio of the timescales governing
the various cooling processes to the timescale for pion decay. Here
$\alpha = (1/\sqrt{L_{52}})(\Gamma_{j}^{(L)}/100)$.
The total cooling timescale is $\tau_{\pi-c}=(\tau_{\pi-N}^{-1}
+\tau_{\pi-\gamma}^{-1}+\tau_{\pi-e}^{-1})^{-1}$. For the $\Gamma_{rel}=2$
curves, $\tau_{\pi-\gamma}/\tau_{\pi-D}$ is too large too appear on this
graph, and the $\tau_{\pi-N}/\tau_{\pi-D}$ curve exactly overlaps with 
the $\tau_{\pi-e}/\tau_{\pi-D}$ curve.
\label{fig2}}
\end{figure}

In Fig. \ref{fig2} we show the ratios of the timescales for the pion
cooling processes discussed above to the timescale for pion decay. For
this figure we have approximated the electron and nucleon number
densities in the post-shock region as being equal and we have also
made the assumption that $\Gamma_{rel}=\Gamma_{\pi}$. As can be seen
from Fig.  \ref{fig2}, for $\Gamma_{rel}=2$ and $r\gtrsim 1.5 \cdot
10^9 \sqrt{L_{52}}(100/\Gamma_j^{(L)}){\rm cm}$, pion cooling
processes are unimportant (e.g. the timescale for all cooling
processes is more than twice the decay timescale and less than 30$\%$
of the pion energy is lost). For $\Gamma_{rel}=10$, pion cooling
processes are unimportant for $r\gtrsim 7\cdot
10^9\sqrt{L_{52}}(100/\Gamma_j^{(L)}){\rm cm}$.

On average, neutrinos carry away $\sim 2/3$ of the pion energy. The
typical energies of the three neutrinos the pion (and daughter muon)
decays to are in the range 30-50MeV as measured in the pion rest
frame.  This implies observed neutrino energies in the range $\sim
[1/5-1/3] (E_{\pi}^{(L)}/\xi)(1+z)^{-1}{\rm GeV}$, with $z$ the redshift at
which the burst occurs. In Table \ref{thetable}, typical
neutrino energies for the different simulations of Zhang et al. are
shown. Note that the observed neutrino energy scales linearly with jet
Lorentz factor. For example, for $\Gamma_{j}^{(L)}=400$, the expected
neutrino energies are a factor of approximately four higher than shown in
Table \ref{thetable}.

The fraction of jet kinetic energy lost to neutrinos is
approximately
\begin{equation}
\label{flost}
(1-Y_e)\left({ 2\over 3}\right) \left({2 \over 3}\right)
{E_{\pi}^{(L)} \over \Gamma_j^{(L)} m_N c^2}
\end{equation}
 for collisions ocurring at radii large enough that the charged pions
do not cool before decaying. Here the electron fraction $Y_e$ is the
ratio of protons to baryons in the flow. Recent studies indicate that
the outgoing jet is likely neutron rich, with plausible values of
$Y_e$ in the range $1/20\lesssim Y_e \lesssim 1/2$
\citep{pru02,bel02}.  In Eq. \ref{flost}, one factor of $2/3$
approximately accounts for the fraction of charged pions (as opposed
to $\pi^0$'s) produced, while the other factor of $2/3$ accounts for
the fraction of the charged pion energy going to neutrinos.  The
estimate in Eq. \ref{flost} assumes that after one inelastic collision
the decelerating neutron does not undergo subsequent inelastic
collisions. For this reason, Eq. \ref{flost} represents a lower bound
to the fraction of jet energy lost to neutrinos.  The true fraction
could be as much as a factor of $\sim 2$ higher, depending on the jet
composition and details of the shocking. For the jet dynamics
calculated by Zhang et al., estimates of the fraction of jet energy
going to neutrinos lie in the range $[0.1-0.2](1-Y_e)$.

To facilitate comparison with studies of other GRB related neutrino
production processes, we closely follow the analysis and notation of
\cite{bah00} and \cite{mes00} in estimating the neutrino detection
rate. In the notation of those authors, the average neutrino detection
rate of events at redshift $z$ is $R_{\nu}\approx (N_t/4\pi
D^2)R_bN_n\sigma_{\nu \bar{\nu}}$. Here $N_t$ is the number of protons
in the terrestrial detector, $D$ is the proper distance out to
redshift $z$, $R_b$ is the GRB rate within a Hubble radius, and $N_n$
is the isotropic equivalant number of neutrons undergoing inelastic
scattering.  The neutrino detection cross section averaged over
neutrinos and anti-neutrinos is $\sigma_{\nu \bar{\nu}}\approx 0.5
\cdot 10^{-38}(\sum E_{\nu} /1 {\rm GeV}) (1+z)^{-1}{\rm cm^2}$
\citep{gai}, where $\sum E_{\nu} \approx (2/3) E_{\pi}^{(L)}$ is the
total energy of all neutrinos resulting from an inelastic
neutron-nucleon collision. Using Eq. \ref{epil} and adopting an
Einstein-de Sitter cosmology with Hubble constant $H=65h_{65}{\rm
km/s/Mpc}$ gives
\begin{equation}
\label{detectionrate}
R_{\nu}\approx 2E_{53} (1-Y_e)\left( {100 \over \Gamma_{j}^{(L)}}
\right)f_{shock} \left({ E_{\pi}^{(L)} \over
m_N c^2}\right) \left( {N_t \over 10^{39}} \right)
\left({ R_b \over 10^3}\right) h_{65}^2 \left({2-\sqrt{2}\over 1+z-\sqrt{1+z}}
\right)^2 {\rm year^{-1}}.
\end{equation}
Here $E_{53}$ is the isotropic equivalant jet energy in units of
$10^{53}{\rm erg}$, and $f_{shock}$ is the fraction of the jet
undergoing a relativistic slowing at radii large enough the produced
pions do not cool and decay to unobservable neutrinos ($r\gtrsim
10^{9}{\rm cm}$ for mildly relativistic collisions and $r\gtrsim
10^{10}{\rm cm}$ for ultra-relativistic collisions).  For
illustration, the models of Zhang et al. predict neutrino detection
rates of $\sim 21/{\rm year}$ (JA), $\sim 7/{\rm year}$ (JB), and
$\sim 10/{\rm year}$ (JC), for $E_{53}=1$, $Y_e=1/2$, and
$f_{shock}$=1/2. For model JB, where the relative Lorentz factors
across the shock deep within the star are very similar to the relative
Lorentz factors typical of internal shocks occurring at $\sim
10^{11}{\rm cm}$, the expected detection rate is roughly the same as
that calculated by \cite{mes00} for neutrinos from internal
shocks. For models JA and JC, in which the jet is more dramatically
slowed, the expected detection rates are somewhat larger, approaching
the sorts of detection rates expected for efficient sideways diffusion
of neutrons into the jet.

As can be seen from Eq. \ref{detectionrate}, $R_{\nu}$ does not depend 
explicitly on the beaming angle of emitted neutrinos. However, as we discuss
below, a temporal and directional correlation between detected neutrinos
and the parent GRB is needed in order to distinguish GRB neutrinos from
background atmospheric neutrinos. If the beaming angle of emitted neutrinos
($\Omega_{\nu}$) is smaller than the beaming angle of GRB photons 
($\Omega_{\gamma}$), then a parent GRB is always visible in coincidence
with detected GRB neutrinos. If the opposite case holds ($\Omega_{\nu}>
\Omega_{\gamma}$), then those neutrinos emitted outside of the photon
beaming angle are effectively lost to observation and the detection rate
is effectively decreased by a factor of $\Omega_{\gamma}/\Omega_{\nu}$. 
Two competing processes affect $\Omega_{\nu}/\Omega_{\gamma}$: the focusing
of an initially wide jet as the jet propagates through the star 
\citep{aloy,zha02}, and the expansion of the high enthalpy jet as it breaks
free of the star \citep{zha02}. Jet focusing tends to increase $\Omega_{\nu}
/\Omega_{\gamma}$ because it makes the jet opening angle within the star 
larger than the jet opening angle at the stellar surface. Expansion of the 
high enthalpy jet tends to decrease $\Omega_{\nu}/\Omega_{\gamma}$. Present
2-d simulations of jet propagation do not definitively determine which process
is dominant, though they are roughly consistent with 
$(\Omega_{\nu}/\Omega_{\gamma})\approx 1$.

It should be emphasized that Eq. \ref{detectionrate} and the detection
rates estimated for GeV neutrinos from GRBs neglect many of the
experimental difficulties associated with detecting these
neutrinos. As noted in the introduction, the detector phototube
spacing must be small enough to allow detection of $\sim 10{\rm GeV}$
muons with $\sim 50{\rm m}$ pathlengths (e.g. see \cite{halz2}). In
addition, the angle of the incident neutrino relative to the GRB must
be sufficiently well known to distinguish a GRB neutrino from the
background atmospheric neutrinos. Rough estimates can be made by
scaling detection rates for Super-Kamiokande to the larger detectors
discussed here. In the search through the Super-Kamiokande data for
GRB neutrinos, a high-energy neutrino ($E_{\nu}>200 {\rm MeV}$)
background rate of $2\cdot 10^{-3}{(20{\rm \, sec})^{-1}}$ was assumed
\citep{kamio}.  Estimating that roughly 1/5 of these have energies
larger than a few GeV, approximating the volume of a km$^3$ detector
as being $\sim 10^5/5$ times larger than the volume of
Super-Kamiokande, and supposing the angle of the incident neutrino
relative to the GRB position is known to within a cone of $\sim
10\degr$ half-opening angle gives a background in a km$^3$ detector of
$\approx 0.1({\rm 20 \, sec})^{-1}$.  With this background rate, a total
sample of 1000 GRBs, and no other information, a detection of 20
neutrinos from 20 separate GRBs (out of a thousand) is not
statistically significant, but the detection of two or three neutrinos
from a single GRB is quite significant (see Table 1 in Fukuda et al.
for detailed estimates). As a definite example of the potential
of large neutrino detectors, the slowing of a relativistic jet in GRB
980425 ($z=0.0085$ \citep{galama}) should have given roughly 100
neutrino detections in a km$^3$ detector. Even without directional
information, such a detection would be significant.

\section{Summary}

Numerical simulations of the propagation of a relativistic jet through
a stellar envelope indicate that, deep within the star, instabilities
along the jet-stellar wall interface slow the jet. This slowing gives
rise to a variation in the Lorentz factor of an initially uniform
jet. As we have shown, the slowing of neutrons within the jet also
gives rise to an interesting and potentially detectable neutrino
signal.

Neutrinos produced during the slowing of the jet can precede GRB
photons by tens to several tens of seconds, depending on the jet
dynamics and stellar progenitor. By contrast, neutrinos produced in a
jet that is not subsequently dramatically slowed arrive within $\sim
0.3 (r_{\gamma}/10^{12}{\rm cm})(100/\Gamma_{j}^{(L)})$ seconds of the
GRB photons, where $r_{\gamma}$ is the radius where the jet shocks and
produces observed photons.  The observation of GeV neutrinos preceding
a GRB flash by $\sim 20$ seconds would be an indication that a
relativistic jet had been slowed while optically thick. In turn, this
could imply that the jet had punched through a stellar envelope.

Slowing of the jet within the star also has an impact on the jet
composition.  This is because neutron-neutron collisions, which
dominate the slowing of neutrons if the jet is neutron rich,
preferentially produce protons. Near threshold, for example, roughly
$75\%$ of the inelastic nn(pp) cross section goes toward the
production of a p(n) \citep{mcg}. A jet born near the central black
hole with a very low electron fraction will arrive at large radii with
$Y_e\sim 1/2$. This will influence subsequent nucleosynthesis in the
jet \citep{pgf02,lem02}, as well as the production of neutrinos by
inelastic nuclear processes at large radii.

Apart from composition effects, though, the source of GeV neutrinos we
have discussed arises independently of other proposed sources of GeV
neutrinos in relativistic fireballs. For example, neutron diffusion in
internal shocks at $r > 10^{11}{\rm cm}$ or the transverse diffusion
of neutrons into the jet at larger radii \citep{mes00}, and inelastic
nuclear collisions during neutron decoupling \citep{bah00}, can still occur
and produce neutrinos. In the most optimistic case where all of these
various processes occur, the detection rate of neutrinos in a km-scale
detector could be as high as 30 per year.

\acknowledgments

The author acknowledges helpful correspondence with Weiqun Zhang
regarding the conditions in collapsar jets and several useful suggestions
from the referee John Beacom.  This research has been
supported by the DOE Program for Scientific Discovery through Advanced
Computing (SciDAC; DE-FC02-01ER41176). This work was performed under
the auspices of the U.S. Department of Energy by University of
California Lawrence Livermore Laboratory under contract W-7405-ENG-48.

\clearpage

\begin{deluxetable}{ccccc}
\tablecaption{Estimates of properties of the neutrino signal from
jets with different initial conditions propagating through a star.
\label{thetable}}
\tablewidth{0pt} \tablehead{ \colhead{Model\tablenotemark{a}} &
\colhead{$\Gamma_{rel}$\tablenotemark{b}} &
\colhead{$E_{\nu}(1+z)$\tablenotemark{c}} &
\colhead{$R_{\nu}$\tablenotemark{d}}} \startdata JA & $\sim 6-50$ &
2-3 & 21 \\ JB & $\sim2 - 3$ & 5-8 & 7 \\ JC & $\sim 4 - 9$ & 2-3 & 10
\\ \tablenotetext{a}{Models refer to the different models calculated
in \cite{zha02}} \tablenotetext{b}{Typical Lorentz factors across the
shocks slowing the jet within the star from the simulations of Zhang
et al.}  \tablenotetext{c}{Our calculation of the typical range of
energy in GeV of neutrinos created in the shock slowing the jet as
seen by an observer at rest with respect to the star.}
\tablenotetext{d}{Our estimate of the annual detection rate of
neutrinos with energies of a few GeV in a km-scale detector. In this
estimate, $z=1$,  $Y_e=1/2$, $E=10^{53} {\rm erg}$, and $f_{shock}=1/2$ are
assumed.}  \enddata
\end{deluxetable}

\clearpage


\begin{thebibliography}{}

\bibitem[Aloy et al.(2000)]{aloy}
Aloy, M.~A., M\"uller, E., Ib\'a\~nez, J.~M., Mart\'i, J.~M., 
\& MacFadyen, A.~I. 2000, \apjl, 531, L119.

\bibitem[Bahcall \& M\'esz\'aros(2000)]{bah00}
Bahcall, J.~N. \& M\'esz\'aros, P. 2000, \prl, 85, 1362

\bibitem[Beloborodov(2002)]{bel02}
Beloborodov, A. M. 2002, astro-ph/0209228

\bibitem[Cercignani \& Majorana(1988)]{cer88}
Cercignani, C. \& Majorana, A. 1988, Physics of Fluids, 31, 1064

\bibitem[Chapline \& Weaver(1979)]{cha79}
Chapline, G.F. \& Weaver, T.A. 1979, Physics of Fluids, 22, 1884

\bibitem[Ensman \& Burrows(1992)]{ens92}
Ensman, L. \& Burrows, A. 1992, \apj, 392, 742

\bibitem[Frail et al.(2001)]{frail}
Frail, D.A. et al. 2001, \apjl, 562, L55

\bibitem[Fukuda et al.(2002)]{kamio}
Fukuda, S. et al. (the Super-Kamiokande collaboration) 2002, \apj, 578, 317

\bibitem[Galama et al.(1998)]{galama}
Galama et al. 1998, Nature, 395, 670

\bibitem[Hagiwara et al.(2002)]{pdg}
Hagiwara et al. (Particle Data Group) 2002, Phys. Rev. D, 66, 010001.

\bibitem[Gaisser(1990)]{gai}
Gaisser, T.~K. 1990, Cosmic Rays and Particle Physics 
(Cambridge: Cambridge Univ. Press)

\bibitem[Groom et al.(2000)]{groom}
Groom, D.E. et al. 2000, European Physical Journal, {\rm C15}, 1


\bibitem[Guetta \& Granot(2002)]{gue02}
Guetta, D. \& Granot, J. 2002, astro-ph/0212045

\bibitem[Halzen(1999)]{hal99}
Halzen, F. 1999, astro-ph/9908138

\bibitem[Halzen \& Hooper(2002)]{halz2}
Halzen, F. \& Hooper, D. 2002, Rept. Progr. Phys. 65, 1025

\bibitem[Karle(2002)]{bigice}
Karle, A. for the \ice collaboration 2002, astro-ph/0209556

\bibitem[Lithwick \& Sari(2001)]{lith}
Lithwick, Y. \& Sari, R. 2001, \apj, 555, 540

\bibitem[Lemoine(2002)]{lem02}
Lemoine, M. 2002, \aap, 390, L31

\bibitem[MacFadyen \& Woosley(1999)]{mac99}
MacFadyen, A. I., \& Woosley, S. E. 1999, \apj, 524, 262

\bibitem[McGill et al.(1984)]{mcg}
McGill, J.~A. et al. 1984, Phys. Rev. C, 29, 204
\bibitem[Mannheim \& Schlickeiser(1994)]{man94}
Mannheim, K. \& Schlickeiser, R. 1994, \aap, 286, 983

\bibitem[M\'esz\'aros \& Rees(2000)]{mes00}
M\'esz\'aros, P. \& Rees, M.J. 2000, \apjl, 541, L5

\bibitem[M\'esz\'aros \& Waxman(2001)]{mes01}
M\'esz\'aros, P. \& Waxman, E. 2001, \prl, 87, 1102

\bibitem[Mott-Smith(1951)]{mot51}
Mott-Smith, H.M. 1951, Physical Review, 82, 885

\bibitem[Piran, Shemi, \& Narayan(1993)]{coast}
Piran, T., Shemi, A., \& Narayan, R. 1993, \mnras, 263, 861

\bibitem[Piran(1999)]{pir99}
Piran, T. 1999, \physrep, 314, 575


\bibitem[Pruet, Guiles, \& Fuller(2002)]{pgf02}
Pruet, J., Guiles, S., \& Fuller, G.~M. 2002, \apj, 580, 368

\bibitem[Pruet, Woosley, \& Hoffman(2002)]{pru02}
Pruet, J., Woosley, S.E., \& Hoffman, R.D. 2002, \apj, in press


\bibitem[Vietri \& Stella (1998)]{vie98}
Vietri, M., \& Stella, L. 1998, \apjl, 507, L45

\bibitem[Vietri \& Stella (1999)]{vie99}
Vietri, M., \& Stella, L. 1999, \apjl, 527, L43

\bibitem[Waxman(1995)]{wax95}
Waxman, E. 1995, \prl, 75, 386

\bibitem[Waxman \& Bahcall(1997)]{wax97}
Waxman, E. \& Bahcall, J. 1997, \prl, 78, 2292

\bibitem[Woosley (1993)]{woo93}
Woosley, S. E. 1993, \apj, 405, 273

\bibitem[Zhang, Woosley, \& MacFadyen (2003)]{zha02}
Zhang, W., Woosley, S. E., \& MacFadyen, A.I. 2003, \apj, 586, 356


\end{thebibliography}
\end{document}